# The behavior of magnetic ordering and the Kondo effect in the alloys, $Ce_2Rh_{1-x}Co_xSi_3$: Evidence for Fermi-surface change from bulk data during magnetic ordering-QCP transformation and applicability of SDW picture

Swapnil Patil, Kartik K Iyer, K. Maiti, and E.V. Sampathkumaran[*]

*Department of Condensed Matter Physics and Materials Science, Tata Institute of Fundamental Research, Homi Bhabha Road, Colaba, Mumbai 400005, India*

The results of magnetic susceptibility, electrical resistivity ($\rho$), and heat-capacity measurements as a function of temperature are reported for the alloys, $Ce_2Rh_{1-x}Co_xSi_3$, crystallizing in an $AlB_2$-derived hexagonal structure. $Ce_2RhSi_3$ exhibits antiferromagnetic ordering at 7 K. The Néel temperature decreases gradually with the increase in Co concentration. For $x \geq 0.6$, no magnetic ordering is observed down to 0.5 K. Interestingly, the $x= 0.6$ alloy exhibits signatures of non-Fermi liquid behavior, while the Co end member is a Fermi liquid. Thus, a transformation of magnetic-ordering state to non-magnetism via non-Fermi liquid state by isoelectronic chemical doping is evident in this solid solution. The heat capacity data reveal dominant role of Kondo compensation in the magnetically ordered state. The electrical resistivity data for $x= 0.2$ and 0.3 alloys show an upturn at respective Néel temperatures, establishing the formation of a magnetism-induced pseudo-gap for these intermediate compositions alone as though there is a gradual Fermi surface transformation as the quantum critical point is approached.

## I. INTRODUCTION

The investigation of competition between the Kondo effect and magnetic ordering in f-electron systems continues to be a topic of active experimental as well as theoretical research [See, for instance, Refs. 1-4 and references cited therein]. In recent years, the studies on the ternary Ce compounds, belonging to $AlB_2$-derived hexagonal structure, have been initiated and interesting features in the magnetic phase diagram have been found. For example, while the stoichiometric compounds, $Ce_2AuSi_3$ and $Ce_2PdSi_3$, are found to behave like cluster-spin-glasses at 3 and 3.5 K respectively, gradual replacement of Au/Pd by Co induces long range antiferromagnetic ordering; the magnetic-non-magnetic phase boundary appears around $x= 0.7$ in these pseudoternary alloys, $Ce_2(Au,Pd)_{1-x}Co_xSi_3$ [5,6]. The Doniach's magnetic phase diagram [7] (as a function of unit-cell volume) for these alloys is different from what was known earlier in the literature for f-electron systems. This led to significant theoretical advancement in understanding the competition between magnetism and the Kondo effect [8]. In sharp contrast to the glassy behavior of Au and Pd systems, the Rh analogue, $Ce_2RhSi_3$, has been found to undergo long range antiferromagnetic ordering at $(T_N=)$ 7 K [9]. The influence of lattice compression and expansion by chemical substitutions at the Ce and Si sites as well as by high pressure studies has been investigated [10]. It was inferred that this compound lies nearly at the peak of Doniach's magnetic phase diagram. Therefore, in this ternary family, the pseudoternary solid solution, $Ce_2Rh_{1-x}Co_xSi_3$, serves an ideal system to investigate the competition between magnetism and the Kondo effect based on isoelectronic substitution at the Rh site. Therefore, in this paper, we report the evolution of the properties as a function of $x$ in this solid solution through the measurements of magnetization, electrical resistivity ($\rho$) and heat-capacity (C) We find features attributable to antiferromagnetism → non-Fermi-liquid (NFL) → (non-magnetic) Fermi-liquid with increasing Co content within this solid solution. Interestingly, the signature of Kondo compensation effect is observed in the heat-capacity data even in the magnetically ordered state and Fermi surface seems to change with $x$ at the magnetic transition as indicated by an upturn in $\rho$ for some intermediate compositions.

## II. EXPERIMENTAL DETAILS

The polycrystalline samples, $Ce_2Rh_{1-x}Co_xSi_3$ ($x=$ 0.0, 0.1, 0.2, 0.3, 0.4, 0.6, 0.8 and 1.0), were prepared by arc melting of stoichiometric amounts of constituent elements in an atmosphere of high-purity argon. The molten ingots were subsequently homogenized in vacuum at 800 C for 5 days. X-ray diffraction patterns (Cu $K_\alpha$) reveal that the alloys are single phase forming in a $AlB_2$-derived hexagonal structure.

Magnetic susceptibility ($\chi$) measurements (T= 1.8 – 300 K) in the presence of a magnetic field of 5 kOe as well as isothermal magnetization (M) measurements at 1.8 K were performed employing a commercial vibrating sample magnetometer (Oxford Instruments, U.K.). Temperature dependence of electrical resistivity was tracked (1.8-300 K) by four-probe method employing silver paint for making electrical contacts of the leads with the sample. Heat-capacity data (1.8-20 K) were obtained by a relaxation method with the help of a commercial physical properties measurements system (Quantum Design, USA); for some Co rich compositions, we could take heat-capacity data down to 0.5 K.

## III. RESULTS AND DISCUSSION
### A. X-ray diffraction patterns

In figure 1, we show typical x-ray diffraction patterns, obtained for $x$= 0.0, 0.6 and 1.0 samples. The presence (or the absence) of superstructure lines as shown in an inset of figure 1 below $2\theta < 15°$ is usually taken as an indication for (or against) doubling of *a* and *c* parameters in the literature. On this basis, conclusion on doubling of both *a* and *c* is straightforward for $x \leq 0.2$, whereas, for $x > 0.2$, it may appear that only *a* is doubled. However, these superstructure lines are so weak that it can escape detection. A careful analysis of the patterns indicates that two weak peaks – indexed (112) and (103) - appearing around the most intense line is also a characteristic of doubling of both the cell parameters. These weak peaks appear for $x > 0.2$ as well and therefore we conclude that the superstructure is retained across entire solid solution (see table 1 for cell constants). We have also performed scanning electron microscopic studies and we found that the specimens are single phase without any evidence for any other phase with a composition homogeneity as revealed by energy-dispersive x-ray analysis. We further note that x-ray diffraction lines undergo a gradual shift towards higher angles with increasing Co content, without any unusual broadening for intermediate compositions. This is demonstrated in the two insets of figure 1 for few typical compositions for two diffraction lines; it is straightforward to conclude that the intermediate compositions are not mere mixtures of Rh-rich and Co-rich members.

### B. Magnetization

The results of magnetic susceptibility measurements are shown in figures 2 and 3. As reported earlier [9, 10], there is a peak in $\chi(T)$ at 7 K for $x$= 0.0 due to antiferromagnetic ordering (see Fig. 2). For a small substitution of Rh by Co (*viz.*, $x$= 0.1), there is a qualitative change in the nature of the $\chi(T)$ curve below $T_N$ and the magnetic ordering shifts to a lower temperature (to 5.5 K). For a further increase of Co concentration, say, for $x$= 0.2 and 0.3, the magnetic order sets in at about 4.5 and 3 K respectively. For x= 0.4, the tendency of $\chi$ to flatten around 2 K is indicative of the onset of magnetic transition at this temperature. There is no evidence for a feature due to magnetic ordering for other compositions. Thus, there is a gradual decrease of $T_N$ with increasing Co concentration. In the case of $Ce_2Au_{1-x}Co_xSi_3$ [Ref. 5], the ordering temperature goes through a peak as a function of *x*, presumably due to a significantly larger unit-cell volume change across the series as well as due to pronounced changes in the electronic structure following substitution of a 3d element in the place of a 5d element. In this respect, the behavior of Co substitution in $Ce_2PdSi_3$ is similar to that of the present solid solution. The *x*-dependence of $T_N$ is consistent with the proposal [10] that this Rh compound lies at the peak of Doniach's magnetic phase diagram.

In the paramagnetic phase, inverse $\chi$ follows Curie-Weiss behavior above 150 K (see figure 3) and there is a gradual deviation from high temperature linearity as the temperature is lowered further possibly due to crystal-field effects. The sign of the paramagnetic Curie temperature ($\theta_p^h$) in this high temperature range remains negative and the magnitude essentially shows an increasing trend with increasing Co content (see table 1). The $\theta_p$, with its sign being negative and with its magnitude increasing with increasing chemical pressure, has to be related to the Kondo temperature ($T_K = n\theta_p$ where *n* varies from 0.2 to 1.0, see Ref. 11), since $T_N$ shows the reverse trend with increasing chemical pressure. It should be remarked that, if the value, $\theta_p^l$, is determined from the data at lower temperatures, say, in the range 10 to 20K [Ref. 12], then the magnitudes are smaller than $\theta_p^h$ (see the insets in figure 3 and table 1) as though the crystal-field-split doublet ground state is characterized by a lower $T_K$ – a well-recognized fact in the literature [13]. It is to be noted that the change of slope of the plot in figure 3 below 25 K is much steeper for cobalt-rich end when compared with $x<0.6$, thereby revealing a sudden change in the $\theta_p$ for the ground state for $x >0.4$. It may be added that the value of the effective moment ($\mu_{eff}$) obtained from the high temperature linear region in the plot of $\chi^{-1}$ versus T is very close to that expected for trivalent Ce ions for all compositions.

In figure 4, we show the isothermal magnetization behavior at 1.8 K. The parent Rh compound exhibits [9, 10] an upward curvature in the field range of 20 to 30 kOe at 1.8 K. We find (see figure 4) that this spin-reorientation transition disappears for a very small substitution of Rh by Co, that is, for $x=0.1$. Interestingly, for x= 0.2, the upward curvature reappears, but at a higher field (around 60 kOe) compared to that for the parent Rh compound. For the two higher compositions, we do not see this upward curvature, but M rather tends towards flattening. These findings imply that there are subtle changes in the canting of magnetic moment with varying $x$ in this series, somewhat similar to the situation in Au-Co and Pd-Co based solid solutions [5,6]. It may also be remarked that the M-H curves are non-hysteretic (also near zero field) and, in addition, there is no saturation of M in the measured field range; these establish that there is no ferromagnetic component in any of these alloys. On this basis, we rule out spin-glass freezing as well for any of the compositions. In addition, we did not observe a cusp in ac susceptibility for any frequency ( 1 Hz to 1kHz) for any composition, which is against spin-glass freezing; absence of decay of the isothermal remnant magnetization with time at 2 K is also consistent with this conclusion; since these are featureless, these data are not presented.

### C. Electrical resistivity

We now discuss electrical resistivity behavior. Qualitative changes in the shapes of $\rho(T)$ plots with varying composition are evident from figure 5. The double-peaked structure typical of an interplay between the single-ion Kondo and the crystal-field effects has been observed [10] for the parent Rh compound after subtraction of the lattice contribution. This feature is apparent for $x=0.1$ to 0.3 in the raw data itself with a distinct negative $d\rho/dT$ at high temperatures. For a higher Co content ($x=0.8$), the positive $d\rho/dT$ is observed, which represents importance of strong 4f-hybridization [14]. Focusing on the low temperature behavior ($< 50$ K), the features due to onset of magnetic ordering are visible at respective temperatures for $x=0.0, 0.1, 0.2$ and 0.3 (see insets in figure 5). However, for $x=0.2$ and 0.3, the behavior at $T_N$ is markedly different in the sense that there is an upturn (see inset) below $T_N$, rather than a decrease observed for Rh richer compositions. Thus there appears to be differences in the Fermi surface at the onset of magnetic ordering with varying composition. We tend to believe that this upturn signals the formation of spin-density-wave (SDW) and therefore, in these compositions, antiferromagnetism appears to develop via a SDW instability. These observations possibly bear immense relevance to current debates in the literature on an important question whether the Fermi surface transformation evolves gradually (SDW picture) or suddenly ("local magnetism" picture) as one approaches quantum critical point by changing compositions/pressure [15]. For $x=0.3$, there is an additional broad peak around 8 K, attributable to Kondo coherence effects. Thus, $x=0.3$ sample shows an interplay of several phenomena: single-ion Kondo effect and crystal-field effect above 10 K, Kondo-coherence below 10 K, and magnetic ordering induced pseudo-gap at lower temperatures.

For $x=0.4$, the change in d$\rho$/dT around 10 K – well above $T_N$ – is attributable to Kondo coherence effects. Since $T_N$ is close to 2 K for this composition (also see the heat-capacity data below), we could not observe any feature due to magnetic transition. For $x=0.6$, the onset of Kondo coherence effect is shifted to a higher temperatures, viz., around 50 K. The trend observed – that is, gradual upward shift of Kondo coherence temperature - is qualitatively consistent with the by-now well-known fact that the lattice compression increases Kondo interaction strength. This alloy does not order magnetically down to 0.5 K (see the heat-capacity data also below). Below 10 K, the electrical resistivity varies rather linearly with temperature – a behavior which is significantly different from the quadratic temperature dependence expected for Kondo lattices. These findings characterize that this composition is a NFL system. For a further addition of Co, say for $x=0.8$ also, we see a similar linear variation of $\rho$ with T at low temperatures. Significantly, the Co end member, $Ce_2CoSi_3$, exhibits $T^2$-dependence of $\rho$ below 20 K, typical of Fermi liquids. Thus, this solid solution shows characteristic features attributable to magnetic Kondo-lattice → NFL → Fermi-liquid transformation with the $x=0.6$ composition lying at the quantum critical point.

### D.    Heat capacity

In order to get more information about the magnetism of this solid solution, we have performed heat-capacity measurements, the results of which are shown in figures 6a and 6b in different ways in the temperature range of interest (below 10 K). There is no evidence for more than one peak due to magnetic transition in any of the compositions, unlike in Au-Co solid solution [5]. While the peaks appear near respective $T_N$ for $x \leq 0.4$, the intensity of the peak in C(T) keeps decreasing with increasing Co – a behavior markedly different from that in Au-Co and Pd-Co solid solutions [5,6]. A possible scenario is that the 4f electrons tend to become more itinerant with Co substitution, a proposal that could be consistent with SDW idea proposed to explain the upturn in $\rho$(T). In addition, Kondo compensation effect coexisting with magnetic ordering can also suppress the magnetic moment. To address this question, we determined 4f-derived entropy ($S_{4f}$) following the procedure suggested in Ref. 16. For this purpose, we have measured heat-capacity of $La_2RhSi_3$ to serve as a reference for lattice contribution; this is a fairly good assumption for the lattice part for Rh rich compositions, particularly in the temperature range of interest (< 10 K) where the phonon contribution becomes weak. From $S_{4f}$ data thus derived, the value of $T_K$ was empirically estimated [17, 18] as the temperature at which the value of $S_{4f}$ per Ce mol is about $2 \times (0.5R\ ln2)$. The value of $T_K$ thus obtained increases with $x$ (about 12 K . for $x = 0.0$ and 0.1, 20 K for $x= 0.2$, about 30 K for $x= 0.3$). This trend in $T_K$ follows the one in $\theta_p^1$. We therefore conclude that the Kondo compensation effect must be playing a role in reducing the intensity of C peak following gradual replacement of Rh by Co. In the case of Au-Co and Pd-Co solid solutions, the value of $\theta_p^1$ is much smaller for magnetically ordered alloys (<5K) and in some cases the sign is even positive. Therefore, the influence on the C peak by Kondo compensation effects is expected to be small in these latter cases, in qualitative agreement with experimental observations [5, 6]. For $x > 0.4$, the C data confirm that there is no magnetic ordering down to 0.5 K, but for a weak feature in C(T) at 6 K which often arises [18] from traces of magnetic Ce oxide impurities (less than 1%). Another point of emphasis is that C/T varies logarithmically with temperature below 4 K (see Fig. 6b) for $x= 0.6$ – a characteristic feature of non-Fermi liquid systems [19] – supporting the conclusion from the $\rho$ data. There is a weaker upturn in C/T below 2 K for $x=0.8$ as well, but it is so weak that we are not able to ascertain its functional form. For the Co end-member, C/T is nearly constant (about $\gamma= 160$ mJ/mol $K^2$) at low temperatures, confirming again that this compound is a Fermi liquid [20, 21]. In short, the heat-capacity data support the conclusions from the ρ data. A point of note is that the ratio of the $T^2$-coefficient ($A$) of ρ and γ is about $4 \times 10^{-6}$ μΩ·cm·mol$^2$·K$^2$·mJ$^{-2}$, which is a deviation from the originally proposed Kadowaki-Woods relation [22]. Such deviations have not been demonstrated so commonly for Ce systems [23-26] and are attributed to the role of degeneracy on this relationship [27].

## IV. CONCLUSION

The magnetism of $Ce_2Rh_{1-x}Co_xSi_3$ is different in many respects from those of Au-Co and Pd-Co solid solutions in the $AlB_2$-family. Notably, the features attributable to antiferromagnetism to non-magnetism via non-Fermi liquid behavior is found in this solid solution, without complications from spin-glass or any complex phase diagram noted in other solid solutions of this structure. There is a profound suppression of heat-capacity peak at the magnetic transition in the magnetically ordered compositions. It therefore appears that there is an increasing importance of Kondo compensation effect in the magnetically ordered state with increasing Co content, unlike other solid solutions studied to date in this family. For some intermediate compositions, before losing magnetic ordering in this solid solution, the electrical resistivity data indicate the formation of a pseudo-gap at the magnetic transition; this finding could be relevant to current debates [15] on the nature of Fermi surface transformation as one approaches quantum critical point and therefore it is worth subjecting these alloys to more studies. Thus, $AlB_2$-derived family of pseudo-ternary alloys offers an ideal playground for any further advancement in understanding the competition between magnetism and the Kondo effect considering a variety of behavior encountered in the physical properties.

**Table 1:** The lattice constants, *a* and *c*, unit-cell volume (*V*), paramagnetic Curie temperatures at low ($\theta_p^l$) and high ($\theta_p^h$) temperatures (as described in text), and Néel temperature ($T_N$), for the alloys, $Ce_2Rh_{1-x}Co_xSi_3$.

| X | a (Å) ± 0.004 Å | c (Å) ± 0.004 Å | V (Å³) | $\theta_p^l$ (K) ± 1 K | $\theta_p^h$ (K) ± 5 K | $T_N$ (K) |
|---|---|---|---|---|---|---|
| 0.0 | 8.237 | 8.445 | 496.2 | -7 | -65 | 7.0 |
| 0.1 | 8.230 | 8.445 | 495.4 | -7 | -80 | 5.5 |
| 0.2 | 8.214 | 8.441 | 493.2 | -9 | -80 | 4.5 |
| 0.3 | 8.202 | 8.445 | 492.0 | -11 | -95 | 3.0 |
| 0.4 | 8.201 | 8.439 | 491.5 | -14 | -125 | 2.0 |
| 0.6 | 8.159 | 8.428 | 485.9 | -22 | -150 | - |
| 0.8 | 8.109 | 8.416 | 479.2 | -40 | -220 | - |
| 1.0 | 8.098 | 8.402 | 477.1 | -44 | -280 | - |

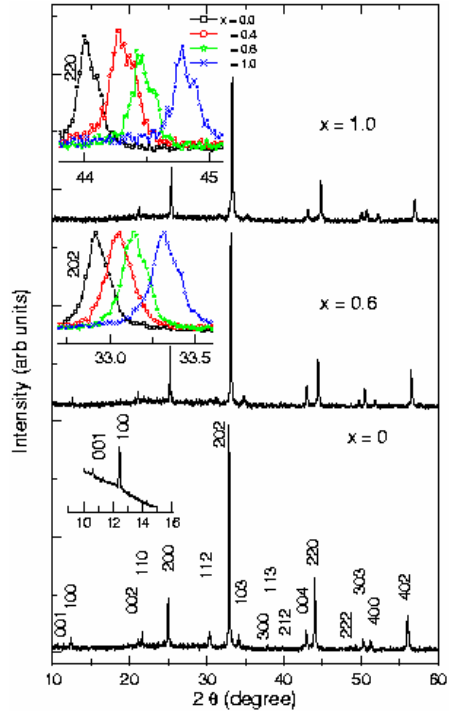

Figure 1:

(color online) X-ray diffraction patterns (Cu $K_\alpha$) for selected compositions (*x*= 0.0, 0.6, and 1.0) in the series, $Ce_2Rh_{1-x}Co_xSi_3$. Miller indices, which is same for all compositions, are also marked. The superstructure lines appearing in low angle side are shown in an expanded form for *x* = 0.0 in an inset. In addition, the diffraction peaks for selected compositions at two different angle-ranges are plotted in an expanded form to highlight that the pattern moves systematically with varying composition (A weak splitting of the lines is due to slightly different wavelengths of $K_{\alpha 1}$ and $K_{\alpha 2}$ x-rays).

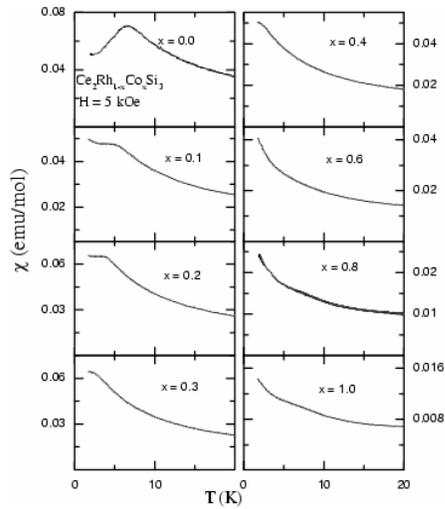

Figure 2:

Magnetic susceptibility ($\chi$) as a function of temperature (T) below 20 K for the alloys, $Ce_2Rh_{1-x}Co_xSi_3$.

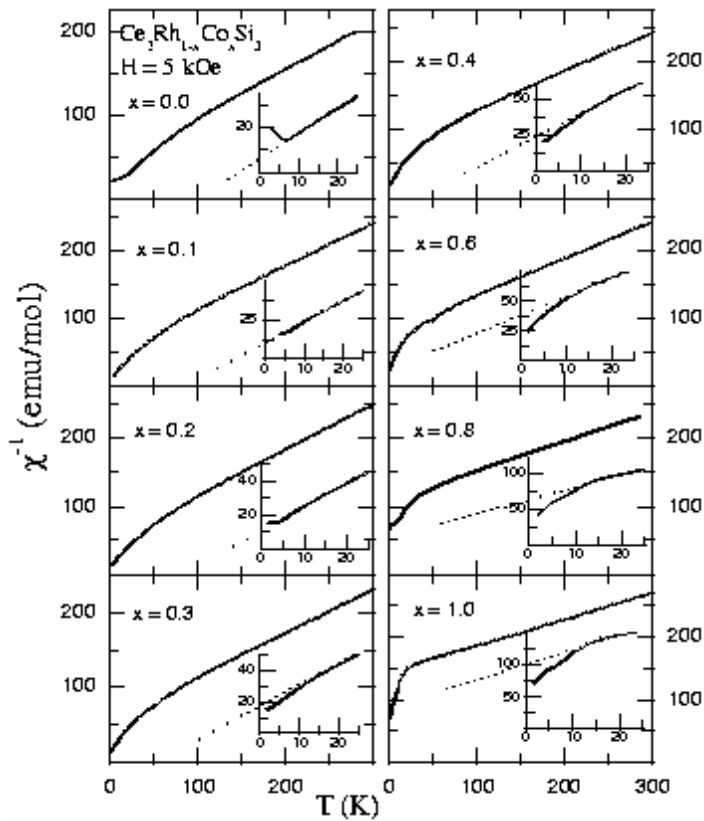

Figure 3:

Inverse susceptibility as a function of temperature for the alloys, $Ce_2Rh_{1-x}Co_xSi_3$. In the inset, the low temperature data are plotted in an expanded form and the dotted lines through the linear region are drawn to show how the paramagnetic Curie temperature at low temperatures varies with composition.

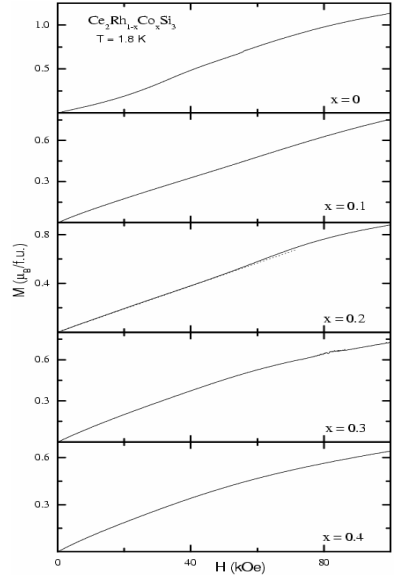

Figure 4

Isothermal magnetization at 1.8 K for $x \leq 0.4$ for $Ce_2Rh_{1-x}Co_xSi_3$. The curves for upward and downward field cycles fall one over the other. A dotted line for $x= 0.2$ is a linear extrapolation from the low-field data to show an upward curvature of the data above 45 kOe.

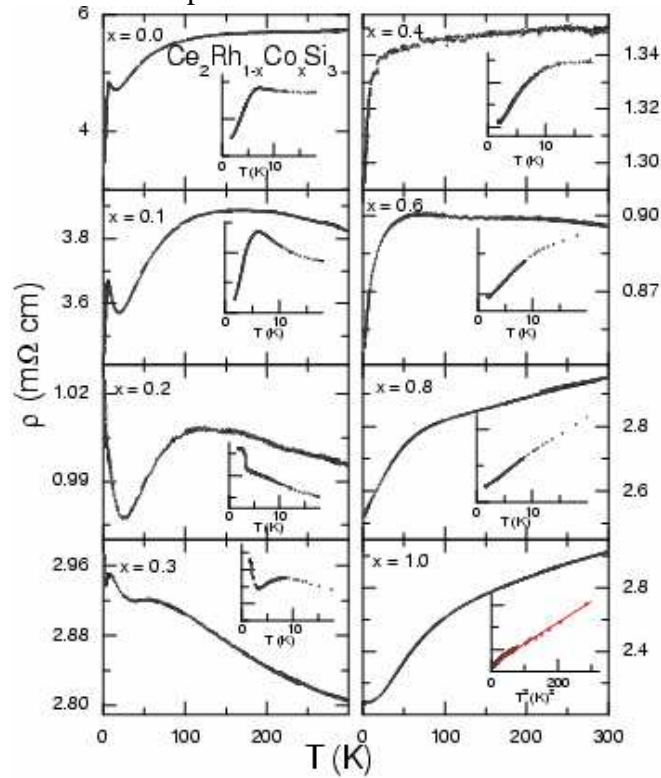

Figure 5:
Electrical resistivity as a function of temperature for the alloys, $Ce_2Rh_{1-x}Co_xSi_3$. The insets show the data in an expanded form; in the inset for $x= 1.0$, a line is drawn through the linear region of the plot.

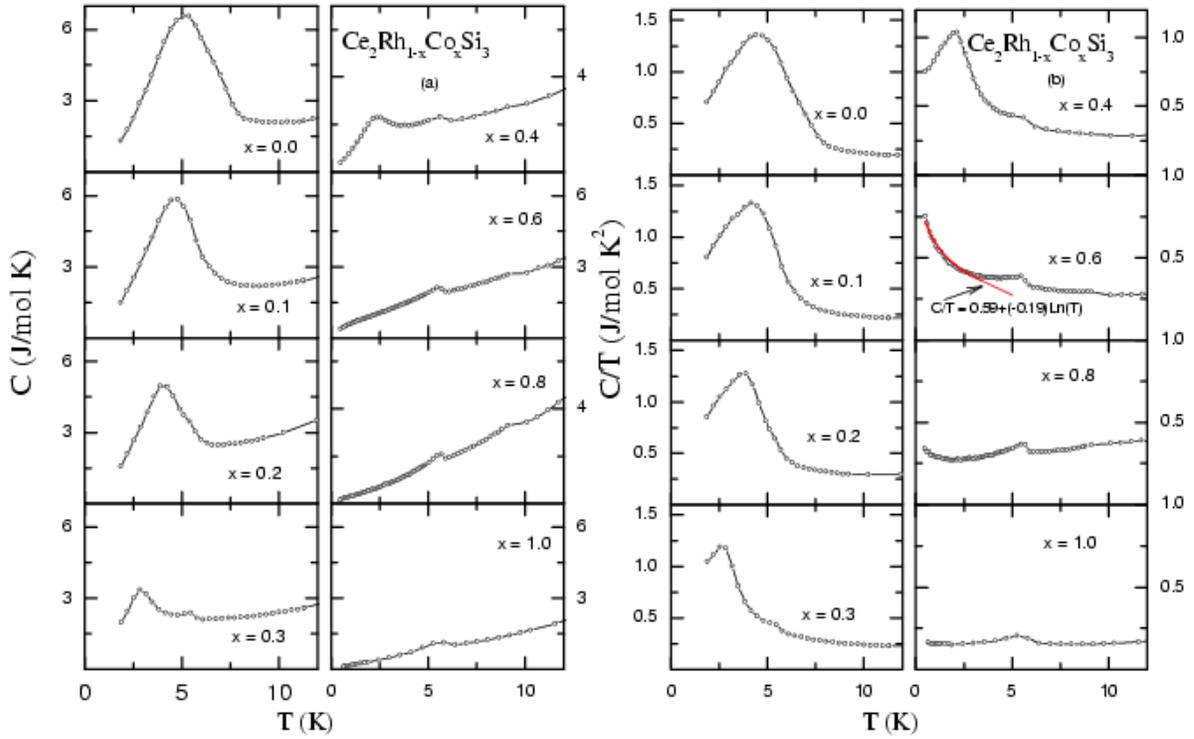

Figure 6:

(a) Heat capacity and (b) Heat-capacity divided by temperature, as a function of temperature for $Ce_2Rh_{1-x}Co_xSi_3$. The lines through the data points serve as guides to the eyes. The C/T data below 4 K for $x= 0.6$ is fitted to a logarithmic variation with temperature.